\documentclass{article}
\usepackage{color}
\usepackage{latexsym}
\usepackage{amsmath}
\usepackage{amsfonts}
\usepackage{amssymb}
\usepackage{indentfirst} 
\numberwithin{equation}{section}
\newcommand{\be}{\begin{equation}}
\newcommand{\ee}{\end{equation}}

\newcommand{\mC }{{\mathbb C}}

\newcommand{\mL}{{\mathcal L}}

\title{The covariance of chiral fermions theory}
\author{K. Andrzejewski$^{1)}$, Y. Brihaye$^{2)}$, C. Gonera$^{1)}$, \\J. Gonera$^{1)}$, P. Kosi\'nski$^{1)}$\footnote{Corresponding author, e-mail: pkosinsk@uni.lodz.pl}, P. Ma\'slanka$^{1)}$ 
\vspace*{1cm}
\\
\small  $^{1)}$Department of  Computer Science,  Faculty of Physics \\   \small and Applied Informatics,
 University of  Lodz, Poland\\
 \small $^{2)}$Physique-Math\'ematique, Universit\'e de Mons-Hainaut,\\
\small  Mons, Belgium
}
\date{}
\begin{document}
\maketitle 
\begin{abstract}
The quasiclassical theory of massless chiral fermions is considered. The effective action is derived using time-dependent variational principle which provides a clear interpretation of relevant canonical variables. As a result their transformation properties under the action of Lorentz group are derived from first principles. 
\end{abstract}
\section{Introduction}
In recent years a renewed interest is observed concerning chiral anomalies, their physical consequences (in particular, for kinetic theory) and relations to topology and Berry curvature \cite{b1}-\cite{b15}. For many purposes (e.g. the dynamics in weak and slowly varying external fields) one can rely on semiclassical approximation. In particular, within this approximation the Weyl equation for helicity $-\frac{1}{2}$ massless charged fermions is replaced by the semiclassical dynamics summarized in the action functional
\be
\label{e1}
I=\int \left((\vec q +\vec A)\cdot d\vec x -(\varepsilon+A^0)dt+\vec\alpha(\vec q)\cdot d\vec q \right),
\ee
where $\vec q$ denotes gauge-invariant (kinetic) momentum, $\vec \alpha(\vec q)$  is the vector potential describing the Berry monopole in momentum space while
\be
\label{e2}
\varepsilon=|\vec q|+\frac{\vec q \cdot \vec B}{2|\vec q|^2}.
\ee
Eq. (\ref{e1})  can be derived from Weyl Hamiltonian by making the semiclassical approximations to path-integral representation of transition amplitude \cite{b10}. The obvious trouble with eq. (\ref{e1}) is that it lacks manifest Lorentz symmetry (in the absence of external fields) or Lorentz covariance, in spite of the fact that it is derived from covariant Weyl equation. In Ref. \cite{b10} a modified transformation rule under Lorentz boosts has been proposed for the particle dynamical variables, involving $O(\hbar)$  corrections, which leaves invariant the dynamics described by the action (\ref{e1}). It reduces to the standard Lorentz form for spinless particles. The modified rules involve helicity dependent terms and close only on-shell. 
\par
The resulting situation attracted much attention \cite{b9,b14}. In particular, Duval et al. were able to derive the modified transformation rules for chiral fermions from Soriau model \cite{b16} of massless spinning particles by applying the procedure called "spin enslaving". Then similar results have been obtained by using the coadjoint orbits method \cite{b15}. 
\par
The present paper is devoted to the further study of the covariance problem for chiral fermions. We consider the latter to be important because it can shed more light on the question of localizability and the meaning of position operator in relativistic quantum theory, in particular in the case of massless particles. In a very nice paper \cite{b18} Skagerstam noticed that the existence of position operator with standard transformation properties and commutation rules would imply that the massless irreducible representation of Poincare group accommodates all $2\lambda+1$ chiralities, which is not the case (note that already Newton and Wigner \cite{b19} found that the position operator with commuting coordinates exists only for $\lambda=0,\frac 12$; in the latter case both helicities should be considered). In fact, the results obtained in \cite{b15} imply that we have the following choice: either (i) the coordinate operator has noncommuting components or (ii) it transforms nonlinearly under rotations. As a result the transformation properties of coordinates of massless particles with nonzero helicity are "exotic" which is nicely intuitively explained in \cite{b10} (see also last Ref. \cite{b9}). This reasoning, when applied to photons, leads to the so-called relativistic Hall effect of light \cite{b20}-\cite{b23}.
The problem has, however, more general context. It is well known \cite{b24} that, due to the fact that in the quantum relativistic regime the presence of interactions excludes particle number conservation, the notion of localizability has a restricted meaning. Obviously, as long as noninteracting theory is considered we do not  have to bother about localizability. Everything can be formulated in momentum space which provides a natural framework for Poincare covariance. One can, of course, also play with enveloping algebra of Poincare one to propose and construct various types of position operators; however, the physical relevance of such activity is rather questionable. The situation changes drastically when interaction is switched on. As beautifully explained by Weinberg \cite{b25}  the most convenient (unique?) way of imposing the Lorentz covariance is to construct Lorentz covariant local fields with the help of Fourier transform supplied by relevant intertwining operators; such fields provide building blocks for constructing covariant interactions. The four coordinates which appear in this way should be considered as parametrizing the Minkowski space-time points. Operationally, they can be defined as positions of some heavy particles; the heavier they are the more unique is the notion of their localizability. The upper bound on their masses is related to the possibility of black hole formation which leads to the well known conclusion that the very notion of continuous space-time breaks at Planck length. In this way we get some feeling concerning the meaning of space-time coordinates. However, the question what is the relation between space-time coordinates and the coordinates of actual particles described by the field under consideration remains unsettled. In the case the interaction is present this problem becomes important. In both nonrelativistic or classical limits our understanding of localizability is quite complete. Moreover, in the weak field regime where particle number is approximately conserved we can look for the effective one-particle dynamics which should be formulated in terms of coordinate and momenta.
 
	The quasiclassical theory of Weyl fermions provides a good laboratory for studying the relation between the dynamical variables characterizing a particle and the corresponding wave function/field. In particular, this concerns the transformation properties of both objects. We are dealing with one-particle approximation to relativistic interacting theory and ask for the transformation rules of the particle dynamical variables following from the transformation properties of the relevant wave function. The transformation rules are derived from first principles because the dynamical variables entering the action are precisely defined. In spite of their form (cf. eq.(2.15)) they are not obtained as a simple extension of nonrelativistic variables but emerge naturally from the relevant variational principle in semiclassical approximation. For example, eq.(2.15) is only valid provided pair creation is neglected. Once the precise definition of dynamical variables in the approximation under consideration is given, their relativistic transformation rules follow from the covariance of Weyl equation (in particular, from usual transformation rules of space-time arguments of fields entering Weyl equation). The advantage of such an approach over the one based on path integral stems from clear interpretation of particle coordinates which allows to replace an educated guess by a straightforward derivation of Lorentz transformation properties.
\par 
The action $I$ (eq. (\ref{e1})) has been derived in \cite{b10} from that for Weyl fermions using the path integral method. However,as it has been pointed out above, in order to provide a clear interpretation of the dynamical variables entering (\ref{e1}) we prefer to use the method popular among condensed matter physicists \cite{b26}-\cite{b29} which is based on time-dependent variational principle.
\par
The paper is organized as follows. In Sec. \ref{s2} we consider free Weyl theory. Interaction with external electromagnetic field is considered in Sec. \ref{s3}. Some conclusions are presented in Sec. \ref{s4}. Appendix is devoted to some technicalities. 
\section{Free Weyl fermions}
\label{s2}
Let us start with free massless left-handed Weyl fermions. The relevant  wave function  obeys Weyl equation
\be
\label{e3}
\sigma^\mu\partial_\mu\phi=0,
\ee 
with $\sigma^\mu=({{ \mathbb I}},-\vec \sigma)$, $\vec \sigma=(\sigma_1,\sigma_2,\sigma_3)$ being Pauli matrices.  Eq. (\ref{e3}) is invariant  under the action of Poincare group given by 
\be
\label{e4}
\phi_\sigma'(x)=(A^\dag)_{\sigma\sigma'}\phi_{\sigma'}(\Lambda x+a),
\ee
where $SL(2,\mC)\ni A\longmapsto \Lambda(A)\in SO(1,3)$ defines  double (universal) covering of the Lorentz group. We shall consider  the single particle theory described  by positive energy  solutions to eq. (\ref{e3}) . They read  
\be
\label{e5}
\phi(\vec x,t)=\frac{1}{(2\pi)^\frac 3 2}\int d^3\vec p c(\vec p)u_+(\vec p)e^{i(\vec p \vec x-|\vec p|t)},
\ee
where $u_+(\vec p)$ is the positive energy  spinor  obeying 
\be
\label{e6}
\vec p \cdot \vec \sigma u_+(\vec p)=-|\vec p|u_+(\vec p),
\ee
together with the normalization condition
\be
\label{e7}
u_+^\dag(\vec p)u_+(\vec p)=1.
\ee

  The above normalization condition is not Lorentz invariant. On the other hand it is convenient because such a combination of spinors enters Lorentz invariant definition of scalar product ( being the space integral of zeroth component of current) and action functional $I$, eq.(\ref{e12}) below (resulting from invariant action functional for Dirac particle in the limit $m\rightarrow0$ with one chirality deleted). Another choice of normalization would only influence the actual form of wave packet profile $c(\vec{p}) $; the normalization condition enters then explicitly, for example, the expression for coordinate. However, all our conclusions remain valid.
Explicitly, $u_+(\vec p)$ can be chosen as 
\label{e8}
\be
u_+(\vec p)=\frac{1}{\sqrt{2|\vec p|(|\vec p|+p^3)}}\left(
\begin{array}{c}
p^1-ip^2\\
-(|\vec p|+p^3)
\end{array}\right).
\ee
The wave function $\phi(\vec x,t)$, given by eq. (\ref{e5}), carries mass zero helicity $-\frac 12$  irreducible  representation of Poincare group; $u_+(\vec p)$ is  the corresponding intertwining operator \cite{b25}. In fact, the generators  of the representation (\ref{e4}) read  
\begin{align}
\label{e9}
\begin{split}
M_{\mu\nu}&=i(x_\mu\partial_\nu-x_\nu\partial_\mu)+m_{\mu\nu},\\
m_{ij}&=\epsilon_{ijk}\frac{\sigma_k}{2}, \quad m_{0k}=\frac{i\sigma _k}{2}.
\end{split}
\end{align}
Their  momentum  counterparts, $\tilde M_{\mu\nu}$, acting  in the space of momentum amplitudes, are given by
\be
\label{e10}
(M_{\mu\nu}\phi)(\vec x,t)=\frac{1}{(2\pi)^\frac 3 2}\int d^3\vec p(\tilde M_{\mu\nu}  c)(\vec p)u_+(\vec p)e^{i(\vec p \vec x-|\vec p|t)}.
\ee
It is straightforward to check that $\tilde M_{\mu\nu}$ take  the standard  form of generators of massless helicity $-\frac 12$ representation  of Poincare  group (actually, its Lorentz subgroup). For example, the boosts read
\begin{align}
\label{e11}
\begin{split}
\tilde M_{01}&=i|\vec p|\frac{\partial}{\partial p^1}+\frac{ip^1}{2|\vec p|}-\frac{p^2}{2(|\vec p|+p^3)},\\
\tilde M_{02}&=i|\vec p|\frac{\partial}{\partial p^2}+\frac{ip^2}{2|\vec p|}+\frac{p^1}{2(|\vec p|+p^3)},\\
\tilde M_{03}&=i|\vec p|\frac{\partial}{\partial p^3}+\frac{ip^3}{2|\vec p|}.
\end{split}
\end{align}
Eqs. (\ref{e11}) coincide with the well-known expressions (see, for example,  \cite{b15,b30}) provided  one takes  into account that the invariant scalar product is defined  here as  $\int d^3\vec p\, \overline{c}_1(p)c_2(p)$ while  usually the explicitly invariant measure $\frac{d^3\vec p}{2|\vec p|}$ is used. Note  that  eqs. (\ref{e11}) are not explicitly  covariant under spatial  rotations; this is because  any choice  of standard vector  for massless representation  of Poincare group necessarily breaks explicit  rotational covariance. Let  us also remind  that eqs. (\ref{e11}) can be obtained  by a straightforward  quantization  of the Hamiltonian  system  invariant under  the action of the Poincare  group, constructed with  the  help of coadjoint  orbit method \cite{b15}. 
\par
Up to now the information contained in eq. (\ref{e3}) (together with the choice of solutions with definite energy sign) is equivalent to that provided by the relevant unitary representation of Poincare group. Therefore, in principle, we do not have to consider space coordinates which are formally introduced by Fourier transform (cf. eq. (\ref{e5})). However, they become important if we are going to describe interactions obeying locality principle necessary to provide relativistic invariance of interacting theory \cite{b25}. The space-time coordinates are commuting c-numbers transforming in the standard way; no exotic terms appear in Lorentz transformation rules. Therefore, the particle coordinates entering the semiclassical action must differ from space coordinates appearing in Weyl equation. In order to find the relation between them we derive the semiclassical action functional following the method based on time-dependent variational principle \cite{b26}-\cite{b29}. It provides a clear interpretation of the dynamical variables entering the action functional. To this end note that the action functional for Weyl fermions, 
\be
\label{e12}
I=i\int d^4x\phi^\dag\sigma^\mu\partial_\mu\phi,
\ee
can be written in the form
\begin{align}
\label{e13}
I&=\int dt d^3\vec x\phi^\dag(\vec x,t)\left(i\frac{\partial}{\partial t}-i\sigma_k\frac{\partial}{\partial x^k}\right)\phi(\vec x,t)
=\int dt\mL(t).
\end{align}
In order to find the effective Lagrangian $\mL(t)$  we put
\be
\label{e14}
\phi(\vec x,t)=\frac{1}{(2\pi)^{\frac 32}}\int d^3\vec p c(\vec p,t)u_+(\vec p)e^{i\vec p \vec x} , 
\ee
where $c(\vec p,t)$ is a wave packet profile. In the spirit of semiclassical approach we assume that $|c(\vec p,t)|$ is strongly peaked at some $\vec p_c(t)$. With this assumption $\mL(t)$ can be easily computed from (\ref{e13}) and (\ref{e14})
\begin{align}
\label{e15}
\mL(t)&=\int d^3\vec x\phi^+(\vec x,t)\left(i\frac{\partial}{\partial t}-i\sigma_k\frac{\partial}{\partial x^k}\right)\phi(\vec x,t)
=\frac{\partial \lambda(\vec p_c,t)}{\partial t}-|\vec p_c|,  	
\end{align}
where $\lambda (\vec p,t)$ is the phase function defined by 
\be
\label{e16}
c(\vec p,t)=|c(\vec p,t)|e^{-i\lambda(\vec p,t)}.
\ee
The first term of the right-hand side  of eq. (\ref{e15})  calls for more transparent interpretation. To this end we  define the ``center of charge'' $\vec x_c$ by
\be
\label{e17}
\vec x_c=\int d^3\vec x\phi^\dag(\vec x,t)\phi(\vec x,t)\vec x,
\ee
(assuming $\phi(\vec x,t)$ is normalized to unity). With our assumption concerning $|c(\vec p,t)|$ eq. (\ref{e17}) yields
\be
\label{e18}
\vec x_c=iu^\dag_+(\vec p_c)\vec\nabla_{p_c}u_+(\vec p_c)+\vec \nabla_{p_c}\lambda(\vec p_c,t). 
\ee
Combining eq. (\ref{e15}) and (\ref{e18}) one finds 
\begin{align}
\label{e19}
\mL(t)&=\vec p_c{\dot{\vec x}_c}-|\vec p_c|-\vec \alpha(\vec p_c)\dot{\vec p}_c+\frac{d}{dt}(\lambda(\vec p_c,t)-\vec x_c\vec p_c) 
\simeq \vec p_c\dot{\vec x}_c-|\vec p_c|-\vec \alpha(\vec p_c)\dot{\vec p}_c,
\end{align}
where 
\be
\label{e20}
\vec \alpha(\vec p)=-iu^\dag_+(\vec p)\vec \nabla_pu_+(\vec p).
\ee
We see that the derivation based on time-dependent variational principle yields indeed a transparent interpretation of dynamical variables entering the effective Lagrangian. Eq. (\ref{e17}) allows us to determine the transformation properties of $\vec x_c$.  Consider, for example, an infinitesimal boost described by $\vec \omega\equiv (\omega^{01},\omega^{02},\omega^{03})$. Eqs. (\ref{e4}) and (\ref{e17}), together with our assumption concerning the semiclassical character of the wave packet, yield the following transformation rule for $\vec x_c$ (valid on-shell, i.e. provided the Weyl equation (\ref{e3}) holds, cf. \cite{b10})
\be
\label{e21}
\delta_0\vec x_c=t\vec \omega-(\vec x_c \cdot \vec\omega)\dot{\vec x}_c-\frac{\vec\omega\times\vec p_c}{2|\vec p_c|^2}+(\vec x_c\cdot\vec \omega)\left(\dot{\vec x}_c-\frac{\vec p_c}{|\vec p_c|}\right) .
\ee
The first two terms on the right-hand side refer to the standard Lorentz transformation rule (the second term accounts for the change of time variable); the third term is the exotic correction proposed in Ref. \cite{b10}. Finally, the last term vanishes on semiclassical "on-shell". Let us compare eq. (\ref{e21}) with the trajectory $\vec x=\vec x(t)$  of classical point particle. Under Lorentz transformation $\vec x'(t')=\vec x(t)+t\vec\omega$, $t'=t+\vec\omega\vec x$ yielding  $\vec x'(t)-\vec x(t)=t\vec\omega-(\vec\omega\vec x)\dot{\vec x}$. This shows that the third term on the right-hand side provides the additional helicity-dependent contribution.
\par 
In order to find the interpretation of eq. (\ref{e21}) in the framework of unitary representation of Poincare group let us consider the question if there exists the coordinate operator corresponding to $\vec x_c$. One is tempted to define
\be
\label{e22}
(\hat{\vec x}_c\phi)(\vec x,t)=\vec x\phi(\vec x,t)=\frac{1}{(2\pi)^\frac32}\int d^ 3\vec p i\vec\nabla_p(c(\vec p,t)u_+(\vec p))e^{i\vec p\vec x}.
\ee
This is, however, wrong because $\vec\nabla_p(c(\vec p,t)u_+(\vec p))$ is not proportional to $u_+(\vec p)$; in other words, $i\vec \nabla_p$ is not a one-particle operator. The cure for this is well-known: as in the case of Newton-Wigner operator \cite{b19}  one has to project on positive energy states. Let us define the projector
\be
\label{e23}
\Pi_+(\vec p)=u_+(\vec p)u_+^\dag(\vec p)=\frac 12\left({ \mathbb I}-\frac{\vec p\vec\sigma}{|\vec p|}\right).
\ee
The particle coordinate operator is now defined as 
\begin{align}
\label{e24}
\begin{split}
(\hat{\vec x}_c\phi)(\vec x,t)&=\frac{1}{(2\pi)^\frac32}\int d^ 3\vec p \Pi_+(\vec p)i\vec\nabla_p\Pi_+(\vec p)(c(\vec p,t)u_+(\vec p))e^{i\vec p\vec x}\\
&=\frac{1}{(2\pi)^\frac32}\int d^ 3\vec p\Pi_+(\vec p) i\vec\nabla_p(c(\vec p,t)u_+(\vec p))e^{i\vec p\vec x}.
\end{split}
\end{align}
Obviously, $\hat {\vec x}_c$ obeys
\be
\label{e25}
\vec x_c=\int d^3\vec x\phi^\dag(\vec x,t)\hat{\vec x}_c\phi(\vec x,t).
\ee 
It is straightforward to define the explicit expression for coordinates  operator in momentum space:
\begin{align}
\label{e26}
\begin{split}
\hat{ x}_c^1&=i\frac{\partial}{\partial p^1}-\frac{p^2}{2|\vec p|(|\vec p|+p^3)},\\
\hat{ x}_c^2&=i\frac{\partial}{\partial p^2}+\frac{p^1}{2|\vec p|(|\vec p|+p^3)},\\
\hat{ x}_c^3&=i\frac{\partial}{\partial p^3}.
\end{split}
\end{align} 
It coincides (modulo the definition of scalar product) with the one given in \cite{b9,b15,b18}. The components of $\hat{\vec x}_c$ obey the commutation rules
\be
\label{e27}
[\hat x_c^k,\hat x_c^l]=\frac{i}{2}\epsilon_{klm}\frac{p^m}{|\vec p|^3}.
\ee
It is not difficult to understand the origin of eq. (\ref{e27}). As it has been already mentioned the standard vector defining massless representation cannot be chosen in rotationally invariant way. As a consequence, for some variables the rotational covariance is implemented in a form of nonlinear realization of $SU(2)$ linearizing on some subgroup $U(1)$. As far as coordinate variables are concerned we are left with the following alternative: either we choose the coordinates transforming in nonstandard way under rotations or they form an usual three-vector but fail to obey canonical commutation rules \cite{b15}. Let us note that in the presence of external fields the commutation rule (2.25) must be generalized. However, the transformation rules for dynamical variables, both in the free (eq.(2.19)) and interacting (eqs.(3.10) and (3.12)) cases can be derived without using general commutation rules so there is no point to quote here the latter.
\par
Passing on shell, $c(\vec p,t)=c(\vec p)\exp{(-i|\vec p|t)}$ one finds the form  of $\vec x_c$ in Heisenberg picture (acting on $c(\vec p)$):
\be
\label{e28}
\hat {\vec  x}_c(t)=\hat{ \vec x}_c+t\frac{\vec p}{|\vec p|}.
\ee
Poincare generators may be expressed in terms of $\vec x_c$ and $\vec p$. For example, the boosts read (cf. eqs. (\ref{e11}) and (\ref{e26})):
\be
\label{e29}
\tilde M_{0k}=\frac 12(|\vec p|\hat x_c^k+\hat x_c^k|\vec p|).
\ee
or 
\be
\label{e30}
\tilde M_{0k}=\frac 12(|\vec p|\hat x_c^k(t)+\hat x_c^k(t)|\vec p|)-p^kt\equiv \tilde M_{0k}(t) -p^kt.
\ee
In order to find the transformation  property of $\vec x_c(t)$  we compute
\begin{align}
\label{e31}
[\hat x_c^l(t),\tilde M_{0k}]&=[\hat x_c^l(t),\tilde M_{0k}(t)-tp^k]
=\frac {i}{2}\left(\frac{p^l}{|\vec p|}\hat x_c^k(t)+\hat x_c^k(t)\frac {p^l}{|\vec p|}\right)
-it\delta_{lk}+\frac {i}{2}\epsilon_{lkm}\frac{p^m}{|\vec p|^2}.
\end{align}
For infinitesimal Lorentz transformation one has 
\be
\label{e32}
\delta \hat{\vec x}_c(t)=[\hat{\vec x}_c(t),i\omega^{0k}M_{0k}].
\ee
Eqs. (\ref{e25}), (\ref{e31}) and (\ref{e32}), together with the assumption concerning the shape of wave packet, imply again the transformation rule (\ref{e21}).
\section{Interaction with electromagnetic field}
\label{s3}
Let us now consider the Weyl fermions interacting with external  electromagnetic field. The Weyl  equation reads  now
\be
\label{e33}
\sigma^\mu(\partial_\mu+iA_\mu)\phi=0,
\ee
which corresponds to the Hamiltonian
\be
\label{e34}
H=\vec \sigma\left(i\frac{\partial}{\partial\vec x}+\vec A\right)+A^0,\qquad \vec A=(A^1,A^2,A^3).
\ee
In order to derive the semiclassical action we follow closely the method presented in \cite{b27}. To this end we consider the wave packet centered at $\vec x_c$ at a given time with its spread small as compared to the length scale of variability of electromagnetic field. The local Hamiltonian (in the terminology of \cite{b27}) reads
\be
\label{e35}
\hat H_c=\vec \sigma(i \frac{\partial}{\partial \vec x}+\vec A(\vec x_c,t))+A^0(\vec x_c,t).
\ee
The solution to local eigenvalue equation 
\be
\label{e36}
\hat H_c\phi_p(\vec x,t)=E_{c_p}\phi_p(\vec x,t),
\ee
 has the form
 \be
 \label{e37}
 \begin{split}
 \phi_p(\vec x,t)&=u_+(\vec p-A(\vec x_c,t))e^{i\vec p\vec x},\\
 E_{c_p}&=|\vec p-\vec A(\vec x_c,t)|+A^0(\vec x_c,t),
 \end{split}
 \ee
 where $u_+(\vec p-\vec A(\vec x_c,t)) $ obeys
 \be
 \label{e38}
 \vec \sigma(\vec p-\vec A(\vec x_c,t))u_+=-|\vec p-\vec A(\vec  x_c,t)|u_+. 
 \ee
 The solution (\ref{e37})  has been chosen as  the one which evolves adiabatically from positive energy solution to free Weyl equation; this choice is dictated by the assumption that we are working in the framework of single-particle theory.
\par
Consider now the wave packet
\be
\label{e39}
\phi(\vec x,t)=\frac{1}{(2\pi )^{\frac 32}}\int d^3\vec p c(\vec p,t)u_+(\vec p-\vec A(\vec x_c,t))e^{i\vec p\vec x},
\ee
and assume again that $c(\vec p,t)$  is strongly peaked at some $\vec p_c(t)$. The consistency condition is that the above wave packet yields the pre-assigned center position which implies that eq. (\ref{e17}) is still valid. The starting point for derivation of effective semiclassical action is again the time-dependent variational principle for the action functional
\be
\label{e40}
I=\int dt\int d^3 \vec x\phi^\dag(\vec x,t)\left(i\frac{\partial}{\partial t}-H\right)\phi(\vec x,t),
\ee
with $H$  an $\phi$  given by eqs. (\ref{e34}) and (\ref{e39}), respectively. Following the steps described  in Ref. \cite{b27} (which involve, in particular, Taylor expansion of potentials to first order) we find after slightly troublesome calculations
\begin{align}
\label{e41}
I&=\int dt\Bigg( (\vec q_c+\vec A(\vec x_c,t))\dot{\vec x}_c-\left(|\vec q_c|+A^0(\vec x_c,t)+\frac{\vec q_c\cdot \vec B(\vec x_c,t)}{2|\vec q_c|^2}\right)-\vec \alpha(\vec q_c)\cdot \dot{\vec q }_c\Bigg),
\end{align}
which agrees with previous  findings; here  $\vec q_c\equiv \vec p_c-\vec A(\vec x_c,t)$  is the  gauge-invariant kinetic momentum. Let us stress again  that the wave  packet method provides a clear  interpretation of the variables entering action  functional. In particular, $\vec x_c$ is again  given by  (\ref{e17}). Its transformation properties  can be derived from eqs. (\ref{e4}) and (\ref{e17}). In particular, the Lorentz  boosts take  the form 
\begin{align}
\label{e42}
\delta_0x^i_c&=\omega^{0i}t-\omega^{0k}x_c^k\dot x_c^i+\omega^{0k}x^k_c\left(\dot x_c^i-\frac{q_c^i}{|\vec q_c|}\right)+\frac{(\vec q_c\times \vec\omega)^i}{2|\vec q_c|^2},
\end{align}
(let us remind that $\vec \omega=(\omega^{01},\omega^{02},\omega^{03}))$.
\par
The first term on the right-hand side is the standard Lorentz transformation rule for the spatial part of $x^\mu$; the second one accounts for the time variation ($\delta t=\omega^{0k}x^k$) since we are working in fixed time (Hamiltonian) formalism while the third term vanishes "on-shell". Therefore, our result agrees with previous findings \cite{b9,b10,b14}. It is important to note that the on-shell condition is defined using the equations of motion to zeroth order in Planck constant, in spite of the fact that we are considering the transformation properties to the first order in $\hbar$. This is because the variation of the action (\ref{e41}) under Lorentz transformations to the order $\hbar^1$ is proportional (up to boundary terms) to the Euler-Lagrange expressions to the order $\hbar^0$  (cf. Ref. \cite{b10}) and one can use the equations of motion to this order to account for time variation (see Appendix). Similar reasoning (although a little bit more involved) gives the transformation rule for the momentum variable. Let us note that
\be
\label{e43}
\vec p_c=\int d^3 \vec p|c(\vec p,t)|^2\vec p=\int d^3\vec x\phi^\dag(\vec x,t)(-i\vec \nabla)\phi(\vec x,t).
\ee
In the noninteracting case eq. (\ref{e43})  implies standard transformation rule for $\vec p_c$. For nonvanishing external fields things get more complicated. The transformation rule for $\vec p_c$ involves fields; moreover, we are interested in transformation properties of gauge-invariant kinetic momentum $\vec q=\vec p-\vec A$ rather than those of canonical momentum $\vec p$. After slightly tedious but straightforward calculations one finds
\begin{align}
\label{e44}
\delta q_c^i&=\omega^{0i}\left(|\vec q_c|+\frac{\vec q_c\cdot\vec B}{2|\vec q_c|^2}\right)+\frac{\left((\vec q_c\times \vec \omega)\times \vec B\right)^i}{2|\vec q_c|^2}+(\vec \omega\cdot \vec x_c)\left[\left(\dot{\vec x}_c-\frac{\vec q_c}{|\vec q_c|}\right)\times \vec B\right]^i\nonumber\\
&+(\omega\cdot \vec x_c)\left[\dot{\vec q}_c-\vec E-(\dot{\vec x}_c\times \vec B)\right]^i .
\end{align}
To derive eq. (\ref{e44}) 
we have used the transformation property of external potentials, ${A'}^\mu(x')={\Lambda^\mu}_\nu A^\nu(x)$, $x'^\mu={\Lambda^\mu}_\nu x^\nu$, corrected by taking into account the last terms on the right-hand of eq. (\ref{e42}).
Using equations of motion (again to the zeroth order in $\hbar$, cf. the remark after eq. (\ref{e42})) one checks that eq. (\ref{e44}) agrees with the previous findings. 
\section{Conclusions}
\label{s4}
  We have considered the dynamics of massless chiral fermions in semiclassical approximation, both for free fermions and fermions interacting with background electromagnetic field. Assuming that the momentum profile of the wave packet is strongly peaked at some value $\vec p_c$ while its space spread is small as compared to the length scale of variability of external field, we were able to derive the effective action for single particle sector (i.e. neglecting pair creation). To this end we used time-dependent variational principle which allows for transparent interpretation of canonical variables $\vec x_c$ and $\vec p_c$. In particular, $\vec x_c$ defines the position of the center of charge of the wave packet (cf. eq. (\ref{e17})). However, even in the free case the commuting space variables $\vec x$ entering the Weyl equation and indexing the space points, cannot be considered as the eigenvalues of coordinate operator defined in the framework of single-particle theory. On the contrary, the properly defined coordinate operator $\hat{ \vec x}_c$ has noncommuting components (cf. eq.  (\ref{e27})) which cannot be diagonalized simultaneously (they rather have to obey $\Delta x^i_c\Delta x_c^j\geq \frac 14\frac{|p^k_c|}{|\vec p_c|^3}$, with $(ijk)$ being a cyclic permutation). The relation between space-time coordinates and the coordinate operator becomes even more complicated in the case of fermions interacting with background electromagnetic field. Once again $\hat {\vec x}_c$ is defined by equations analogous to eqs. (\ref{e22})-(\ref{e24}); however, such a definition involves implicitly the actual value of $\vec  x_c$.
\par
In any case, the above mentioned relation appears to be well defined in the framework of time-dependent variational principle. This allows us to derive the proper transformation rules for particle coordinates and momenta from their very definitions showing the covariance of massless chiral theory (to the first order in $\hbar$). What is more, it provides a deeper insight into the problem of how the space-time arguments entering relativistic wave functions/fields should be interpreted. In particular, it appears that the relation between space-time and particle coordinates depends on the context and the approximation used; the main point is that the interaction should be weak enough to justify neglecting pair creation. Then one can work within one particle approximation and the attributes of single particle acquire well defined meaning.
\par
Finally, let us note that the Weyl equation enjoys larger symmetry group - the conformal one. In fact, an unitary representation of conformal group corresponding to massless particles can be desribed in terms of operators from envelopping algebra of Poincare algebra. Therefore, the results obtained above can be extended to include the scaling and special conformal transformations. Details will be presented elsewhere.
\vspace{0.5cm}
\\
{\bf Acknowledgment}\\
This work has been supported by the grant 2016/23/B/ST2/00727 of National Science Centre, Poland. 
\appendix
\section{Appendix}
\label{sa}
In order to show that one can use the equations of motion to zeroth order in $\hbar$ to account for the time variation, let us  consider the total variation of $\vec x_c$ and $\vec q_c$ obtained by taking into account the full (i.e. as to first order in $\hbar$) equations of motion. Then eqs. (\ref{e42}) and (\ref{e44}) are replaced by 
\be
\label{e45}
\delta x^i_c=\omega^{0i}t+\frac{(\vec q_c\times \vec \omega)^i}{2|\vec q_c|^2}+\vec \omega \cdot \vec x_c\left(\vec \nabla_{  q_c}\left(\frac{\vec q_c\cdot \vec  B}{2|\vec q_c|^2}\right)+\frac{(\vec q_c\times \dot{\vec q}_c)}{2|\vec q_c|^3}\right)^i,
\ee
and
\begin{align}
\label{e46}
\delta q_c^i&=\omega^{0i}\left(|\vec q_c|+\frac{\vec q_c\cdot \vec B}{2|\vec q_c|^2}\right)+\frac{\left((\vec q_c\times \vec \omega)\times \vec B\right)^i}{2|\vec q_c|^2}\nonumber \\
&+(\vec \omega\cdot \vec x_c)\left[\left(\vec \nabla_{ q_c}\left(\frac{\vec q_c\cdot \vec  B}{2|\vec q_c|^2}\right)+\frac{(\vec q_c\times \dot{\vec q}_c)}{2|\vec q_c|^3}\right)\times \vec B\right]^i-(\vec\omega\cdot \vec x_c)\left(\vec \nabla_{  x}\left(\frac{\vec q_c\cdot\vec  B}{2|\vec q_c|^2}\right)\right)^i,
\end{align}
respectively. Calling $I_0$ and $I_1$ the $O(\hbar^0)$ and $O(\hbar^1)$ parts of the action (\ref{e41}) we have to show that $\delta I_0$ vanishes up to boundary terms provided it is computed by taking into account only additional terms appearing on the right-hand sides of eqs. (\ref{e45}) and (\ref{e46}). Computing the relevant variation (cf. Ref. \cite{b10}) one obtains
\begin{align}
\delta I_0&=\int dt\left[(\vec \omega \cdot \vec x_c)\left(\vec \nabla_{  q_c}\left(\frac{\vec q_c\cdot\vec  B}{2|\vec q_c|^2}\right)+\frac{(\vec q_c\times \dot{\vec q}_c)}{2|\vec q_c|^3}\right)\right.
\left.\cdot (-\dot{\vec q}_c+\vec E+\dot{\vec x}_c\times \vec B)\right.\nonumber\\
&+(\vec \omega \cdot\vec x_c)
\cdot\left. \left(\left(\vec \nabla_{  q_c}\left(\frac{\vec q_c\cdot \vec  B}{2|\vec q_c|^2}\right)+\frac{(\vec q_c\times \dot{\vec q}_c)}{2|\vec q_c|^3}\right)\times \vec B\right)\left(\dot{\vec x}_c-\frac{\vec q_c}{|\vec q_c|}\right)\right.\nonumber \\
&\left.-(\vec \omega\cdot \vec x_c)\left(\dot{\vec x}_c-\frac{\vec q_c}{|\vec q_c|}\right)\left(\vec \nabla_{x}\left(\frac{\vec q_c\cdot\vec  B}{2|\vec q_c|^2}\right)\right)+\frac{d(\ldots)}{dt}
\right],
\end{align}
which reduces to the total derivative on-shell (defined by full equations of motion). 
  
\end{document}